%% file: article.tex
\documentclass[12pt]{article}
\usepackage{epsfig}

\input econfmacros.tex
\index{Measurement of heavy-flavour production in pp and Pb--Pb collisions at the LHC with ALICE}

\def\Title#1{\begin{center} {\Large {\bf #1} } \end{center}}

\begin{document}

\Title{Measurement of heavy-flavour production in pp and Pb--Pb collisions at the LHC with ALICE}

\bigskip\bigskip

\begin{raggedright}  

{\it Davide Caffarri, for the ALICE Collaboration \index{Caffarri, D.}\\
INFN sez. di Padova, University of Padova,\
35131 Padova, ITALY \\
European Organization for Nuclear Research (CERN), Geneva, Switzerland
}
\bigskip\bigskip
\end{raggedright}

Latest results on open heavy flavour measurements in pp collisions at 7 TeV and Pb-Pb collisions at $\sqrt{s_{\rm{NN}}} = 2.76~\rm{TeV}$ with the ALICE experiment are presented. The results of the single lepton analyses (electrons at central rapidity, muons at forward rapidity) and D meson reconstruction will be discussed.
 
\section{Introduction}
The main goal of the ALICE experiment is to study the properties of the strongly interacting, high density medium that is expected to be formed in ultrarelativistic heavy ions collisions at the LHC. Heavy quarks are a powerful tool to provide insight in the characteristic of this medium. They are produced in the early stage of the collision via hard scattering and they can interact with the medium via gluon radiation or via collisions with other partons.
In pp collisions, the measurement of the heavy-flavour production cross section allows to test perturbative QCD calculations in the LHC energy regime and they are used as a reference for the study of dense matter effects in Pb--Pb collisions.
The nuclear modification factor ($R_{\rm{AA}}$), obtained by comparing $p_{\rm{T}}$ spectra in pp and Pb--Pb collisions at the same energy, allows to measure the effect of in-medium energy loss.

\section{Heavy flavour measurements in ALICE}
The ALICE experiment is made of three parts: a central barrel in the mid rapidity region $|\eta| < 0.9$, a forward muon spectrometer ($-4<\eta<-2.5$) and a set of trigger and event characterization detectors in the forward  and backward regions~\cite{Aamodt:2008zz}. Thanks to its small material budget and the good transverse momentum resolution, ALICE can track particles from very low up to very high momentum (about $100~\rm{MeV/}\textit{c}$ up to $100~\rm{GeV/}\textit{c}$). Momentum measurement is provided using a solenoid magnet field of 0.5 T, where central barrel detectors are placed. For the analyses that are presented, the relevant ALICE detectors are (from the interaction point to the outer part): the Inner Tracking System (ITS), the Time Projection Chamber (TPC), the Transition Radiation Detector (TRD) for electrons identification, and the Time Of Flight (TOF) to identify hadrons, over the full azimuthal angle. Muons are triggered and reconstructed in the forward muon spectrometer, located downstream of an hadron absorber.  The V0 detector is made of two scintillator hodoscopes, located in the forward and backward rapidity regions; this detector is used for triggering both in pp and Pb--Pb collisions and centrality determination in Pb--Pb. The analysis reported here used a data sample collected with a Minimum Bias trigger, based on the V0 detector and two innermost layers of the ITS equipped with silicon pixels. For muon analyses, reconstructed tracks are required to match with corresponding signals in the muon trigger system. The measurement of the centrality is based on the distribution of signals in the V0 hodoscopes, modeled with a Glauber calculation~\cite{Miller:2007ri}. 

Muons are reconstructed using the 5 tracking stations of the spectrometer ($-4 < \eta < -2.5$). For pp collisions, the heavy flavour muon decay $p_{\rm{T}}$ distribution was obtained by subtracting muons coming from prompt and secondary light hadrons decays from the inclusive spectra. This was not the case for Pb--Pb collisions.
D mesons are reconstructed at mid-rapidity, using their hadronic decays ($\rm{D^{0}} \rightarrow \rm{K} \pi$,  $\rm{D^{+}} \rightarrow \rm{K} \pi \pi$, $\rm{D^{*+}} \rightarrow \rm{D^{0}} \pi$) and exploiting their displaced vertex topology. The Particle IDentification (PID) is the main other selection for this analysis, in order reduce the high combinatorial background. The PID is performed using the TOF information and the specific energy loss signal in the TPC. The prompt charm distributions were determined after subtracting the feed down contribution using FONLL pQCD calculations~\cite{FONLL} and applying detector acceptance and reconstruction efficiency corrections. 
Electrons are identified using TPC and TOF PID up to 6 GeV/c both in pp and Pb--Pb collisions, for the former TRD is also used up to 10 GeV/c. The heavy flavour decay electron spectrum was obtained by subtracting from the inclusive spectra a cocktail of known background sources. The cocktail was built starting from ALICE measured $\pi^{0}~\rm{and}~J/\Psi$ spectrum; other light meson spectra have been added to the cocktail, by scaling the $\pi^{0}$ spectra using the $m_{T}$ scaling. Direct photon spectra were taken from NLO pQCD calculations~\cite{FONLL}.
In pp collisions, beauty decay electrons were studied by applying an impact parameter cut on the tracks in order to select displaced electrons. 

\begin{figure}[!ht]
\centering
\epsfig{file=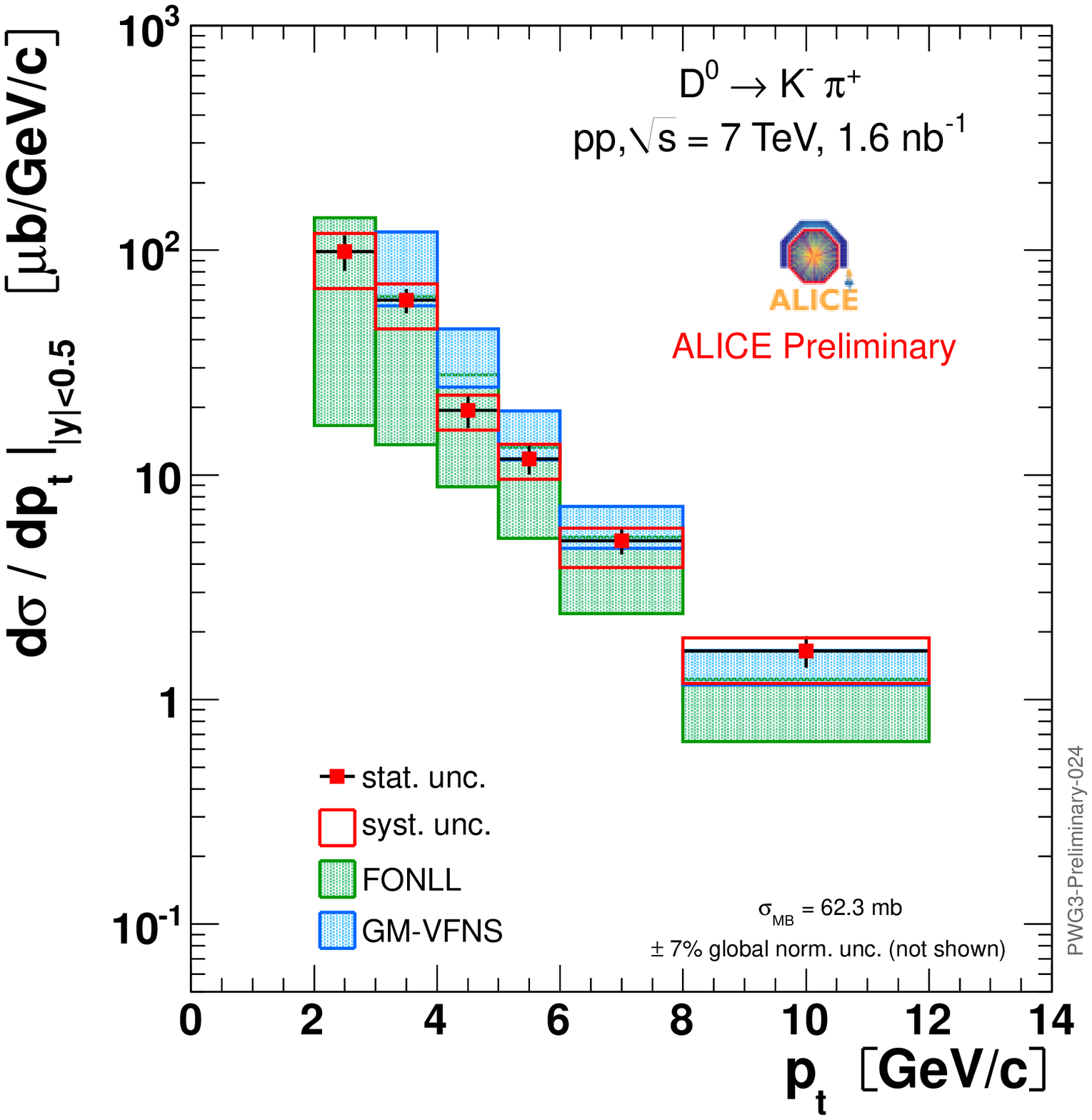,height=2.2in}
\hspace{3mm}
\epsfig{file=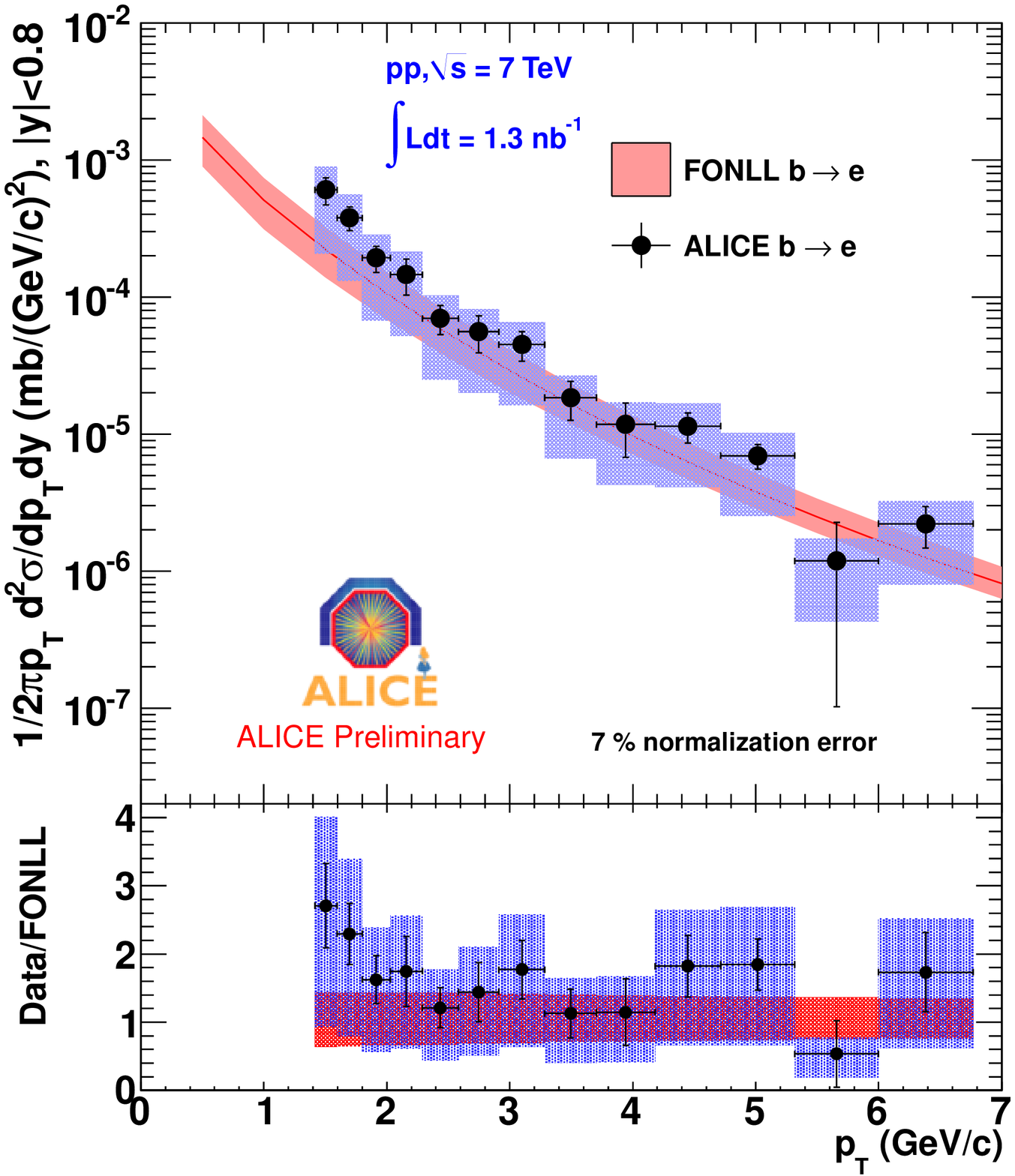,height=2.2in}
\small{\caption{Charm and beauty production cross section in p-p collisions at $\sqrt{s} = 7~\rm{TeV}$ Left: $\rm{D^{0}}$ preliminary cross section compared with FONLL (green) and GM-VFNS (blue) predictions. Right: Cross section for electrons from beauty hadron decays.}
\label{pp_fig}}
\end{figure}

\section{Results}
\begin{figure}[!ht]
\centering
\epsfig{file=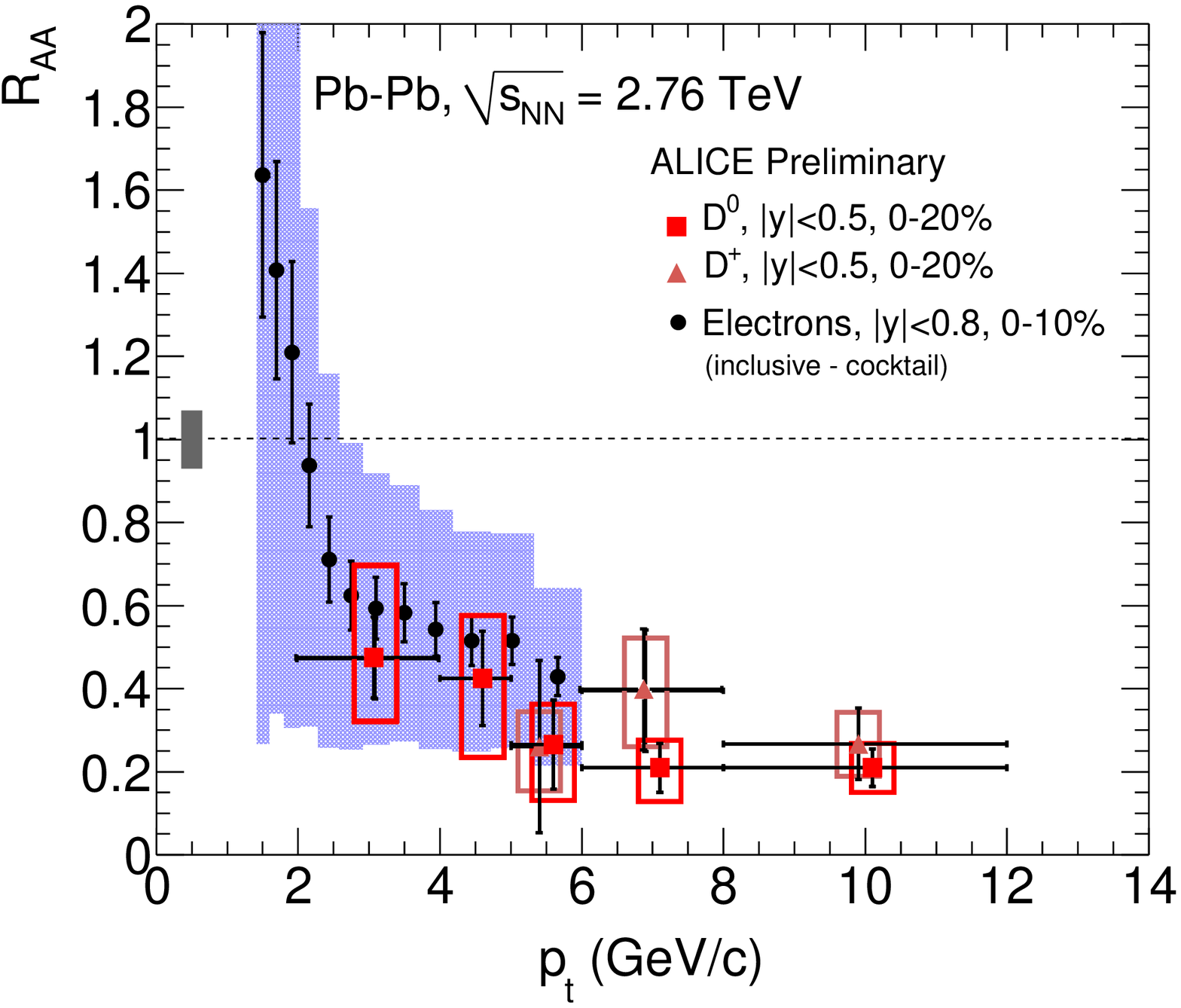,height=2.2in}
\hspace{3mm}
\epsfig{file=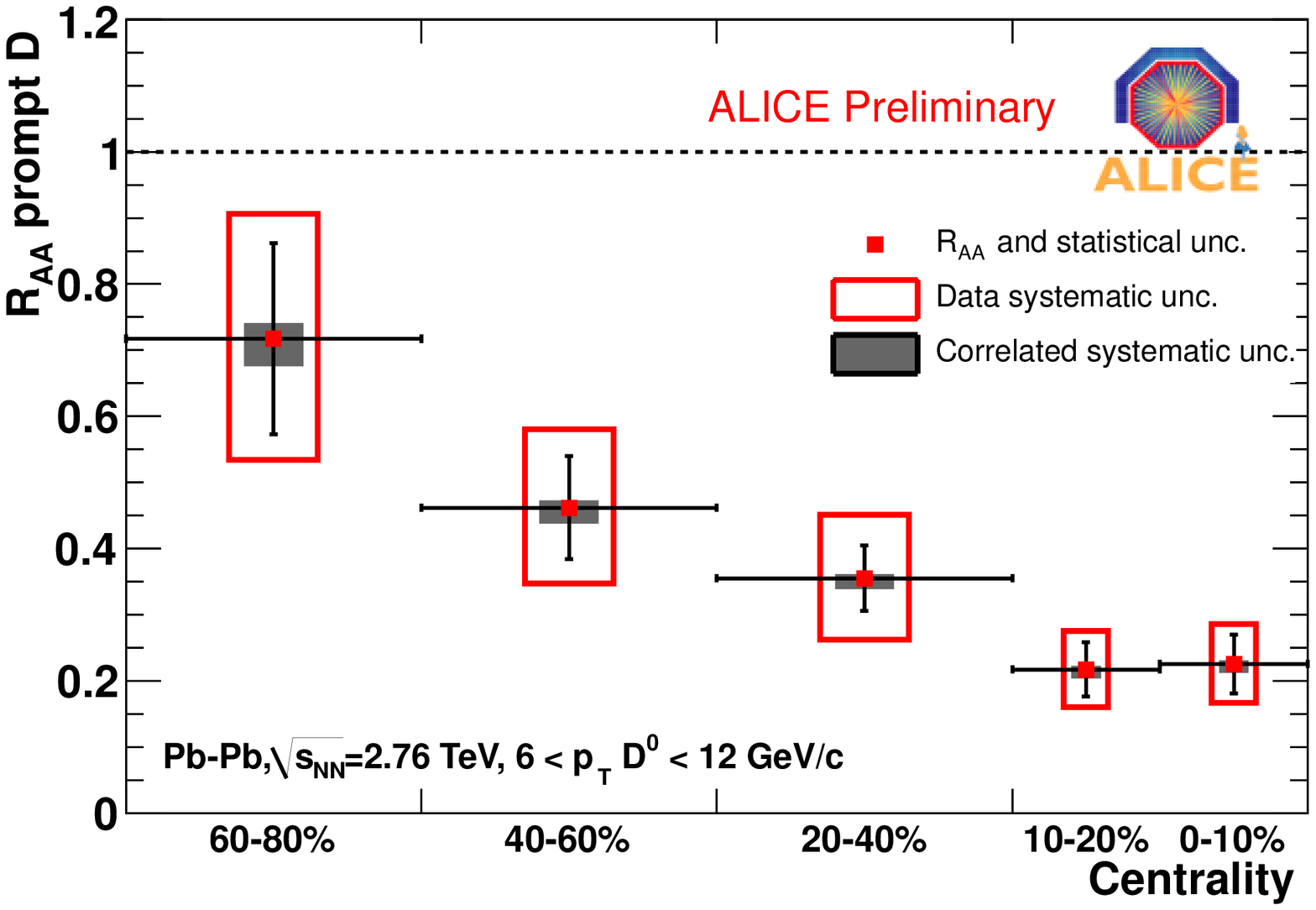,height=2.1in}
\small{\caption{Nuclear modification factor $R_{AA}$ for heavy flavour measured in PbPb collisions at $\sqrt{s_{NN}}= 2.76~\rm{TeV}$ Left: comparison of inclusive - cocktail subtracted electrons and $D^{0}~\rm{and}~D^{+}~R_{AA}$. Right: $D^{0}$ meson $R_{AA}$ centrality dependence for $6 < p_{T} < 12~\rm{GeV/c.}$}
\label{PbPb_fig}}
\end{figure}

The heavy flavour decay muons $p_{\rm{T}}$ and $\eta$ differential cross sections were measured in pp collisions at $\sqrt{s} = 7~\rm{TeV}$~\cite{Dainese:2010ms}, and are well described by FONLL pQCD predictions~\cite{FONLL}. The $\rm{D^{0}}~\rm{,}~D^{+}~ \rm{and}~D^{*+}$ meson cross sections were measured in pp collisions at 7 TeV (Fig.\ref{pp_fig} left) \footnote{Prelminary results are in full agreement with recent final results \cite{ALICEcharm:2011ka}.} and 2.76 TeV, from $2~\rm{GeV/}c$ up to 12 $\rm{GeV/}c$~\cite{Dainese:2010ms}. Also these results agree within uncertainties with FONLL~\cite{FONLL} and GM-VFNS predictions~\cite{GMVFNS}.
The $p_{\rm{T}}$ differential cross section of heavy flavour decay electrons was measured from 0.5 to 10 $\rm{GeV/}c$ and compared to FONLL predictions~\cite{FONLL}. Exploiting the good ALICE impact parameter resolution electrons coming from displaced vertices were selected in order to enhance the beauty contribution within this sample. After subtracting the charm contribution, this analysis allowed to measure the electrons coming from beauty hadron decays (Fig.\ref{pp_fig}, right)~\cite{Dainese:2010ms}.
Open heavy flavour production was also measured in Pb--Pb collisions at  $\sqrt{s_{\rm{NN}}} = 2.76~\rm{TeV}$. To study the effect of the strongly interacting medium formed in Pb--Pb collisions, the ratio of the $p_{\rm{T}}$ spectra in Pb-Pb and pp ($R_{\rm{AA}}$) as well as in central and peripheral collisions ($R_{\rm{cp}}$) are computed:
 
\begin{equation}
R_{\rm{AA}} = \frac{1} {<T_{\rm{AA}}>}  \frac{dN_{\rm{AA}}/dp_{\rm{T}}} {d\sigma_{\rm{pp}}/{dp_{\rm{T}}}}, ~~ R_{\rm{cp}} = \frac{<T_{\rm{AA~periph}}>} {<T_{\rm{AA~central}}>}  \frac{dN_{\rm{AA}}/dp_{\rm{T~central}}} {dN_{\rm{AA}}/{dp_{\rm{T ~periph}}}}
\end{equation}
where $<T_{\rm{AA}}>$ is the average nuclear overlap function, resulting for that centrality class from the Glauber model~\cite{Miller:2007ri}.
In order to obtain the pp reference at $\sqrt{s} = 2.76~\rm{TeV}$, the measurements at $7~\rm{TeV}$ were scaled using FONLL pQCD calculations~\cite{FONLL}. The uncertainty on the scaling factor is mainly obtained by changing the parameters of the predictions (quark mass, renormalization and factorization scales)~\cite{Averbeck:2011ga}. 
For heavy flavour decay muons the comparison between more central (0-10\%) and peripheral (40-80\%) events have been done using $R_{cp}$. A suppression of about a factor 3 was measured in central events with respect to peripheral ones in the $p_{\rm{T}}$ range between 4 and 10 GeV/c~\cite{Dainese:2010ms}. According to the FONLL calculation beauty hadron decays dominate the muon spectrum above 6 $\rm{GeV/}c$~\cite{FONLL}. 
Heavy flavour decay electrons spectrum was measured with the same strategy as in pp collisions. The (inclusive - cocktail) spectrum was measured. The suppression goes from a factor 1.5 to about 4 from 1.5 to 6 GeV/c~\cite{Dainese:2010ms}, but the systematic uncertainties are still quite large, due to difficult background subtraction especially at low $p_{\rm{T}}$.
The $\rm{D^{0}}$ meson spectrum was measured in Pb-Pb collisions from 2 GeV/c up to 12 GeV/\textit{c}, $\rm{D^{+}}$ from 5 to 12 GeV/\textit{c}. The D mesons $R_{\rm{AA}}$ shows a suppression of a factor 4 for $p_{\rm{T}} >5~\rm{GeV/\textit{c}} $ (Fig.\ref{PbPb_fig} left)~\cite{Dainese:2010ms}.  The $\rm{D^{0}}$ nuclear modification factor was also studied as a function of centrality in the $p_{T}$ range between 6 and 12 GeV/c. The suppression is larger for very central events and it starts to decrease going towards more peripheral events (Fig.\ref{PbPb_fig} right). These results indicate a clear energy loss of heavy quarks.

\small{

 }
\end{document}

%% file: econfmacros.tex
\def\beq{\begin{equation}}
\def\eeq#1{\label{#1}\end{equation}}
\def\eeqn{\end{equation}}

\def\beqa{\begin{eqnarray}}
\def\eeqa#1{\label{#1}\end{eqnarray}}
\def\eeqan{\end{eqnarray}}

\let\bar=\overbar

\def\Dslash{\not{\hbox{\kern-4pt $D$}}}
\def\dslash{\not{\hbox{\kern-2pt $\del$}}}

\def\msb{{\bar{\ssstyle M \kern -1pt S}}}

